\documentclass{nature}

\usepackage{graphicx}

\title{Incommensurate magnetic order in the $\alpha$-Fe(Te$_{1-x}$Se$_x$) superconductor systems}

\author{Wei Bao$^{1}$, Y. Qiu$^{2,3}$, Q. Huang$^{2}$, M. A. Green$^{2,3}$, P. Zajdel$^{2,4}$, M. R. Fitzsimmons$^{1}$, M. Zhernenkov$^{1}$, Minghu Fang$^{5,6}$, B. Qian$^{5}$, E.K. Vehstedt$^{5}$, Jinhu Yang$^{6}$, H.M. Pham$^{7}$, L. Spinu$^{7}$ and Z.Q. Mao$^{5}$}

\begin{document}

\maketitle 

\begin{affiliations}
\item Los Alamos National Laboratory, Los Alamos, NM 87545, USA
\item NIST Center for Neutron Research, National Institute of Standards
and Technology, Gaithersburg, MD 20899, USA
\item Department of Materials Science and Engineering,
University of Maryland, College Park, MD 20742, USA
\item Christopher Ingold Building, Department of Chemistry, University College London, UK
\item Department of Physics, Tulane University, New Orleans, LA 70118, USA
\item Department of Physics, Zhejiang University, Hangzhou 310027, China
\item Advanced Materials Research Institute and Department of Physics, University of New Orleans, New Orleans, LA 70148, USA
\end{affiliations}

\begin{abstract}
For high temperature (\mbox{\boldmath$T_C$}) superconductors the formation at low temperature of a condensate, where electrons are paired together to carry the superconducting current, is not disputed. However, the origin of the interaction to cause such pairing is. More than twenty years of research in cuprate superconductors has brought about no consensus, and so the recent discovery of a new group of iron based high temperature superconductors has opened up many new opportunities\cite{Kamihara2008,A033603,A042053,A054630}. Magnetic spin fluctuations is one candidate to produce the bosonic modes that mediate the superconductivity in the ferrous superconductors. Up until now, all of the LaOFeAs and BaFe$_2$As$_2$ structure types\cite{A040795,A062776} have simple commensurate magnetic ground states, as result of nesting Fermi surfaces\cite{A033286,A033426}, with in-plane ($\pi$,0) propagation vectors where the interactions are antiferromagnetic and ferromagnetic along the a- and b- axis, respectively. This type of spin-density-wave (SDW) magnetic order is known to be vulnerable to shifts in the Fermi surface
when electronic densities are altered at the superconducting compositions\cite{A043355}. Superconductivity has more recently been discovered in $\alpha$-Fe(Te,Se), whose electronically active antifluorite planes are isostructural to the FeAs layers found in the previous ferrous superconductors\cite{A072369,A074775,A080474} and share with them the same quasi-two-dimensional electronic structure\cite{A074312}.
Here we report neutron scattering studies that reveal a unique complex incommensurate antiferromagnetic order in the parent compound $\alpha$-FeTe. When the long-range magnetic order is suppressed by the isovalent substitution of Te with Se, short-range correlations survive in the superconducting phase. Therefore, the robust incommensurate antiferromagnetic interactions reported here offer an alternative possibility for mediating the pairing in the ferrous superconductors.

\end{abstract}

The recently discovered ferrous superconductors differ from conventional
phonon-mediated superconductors in an important way: when the nonmagnetic La
in LaFeAs(O,F) is replaced by magnetic lanthanides, $T_C$ increases from 26 K to as high as 55 K\cite{Kamihara2008,A033603,A042053}, in contrast to the
conventional breaking of the $s$-wave Cooper pairs by magnetic ions\cite{mag_sc}. 
The lanthanide oxide/fluoride ``charge reservoir'' layer in the original superconductors\cite{Kamihara2008,A033603,A042053} turns out not to be a requirement for superconductivity and can be replaced by simpler metallic element layers\cite{A054630,A061209,A061301,A064279
}, or completely absent as shown more recently in the $\alpha$-phase of Fe(Se,Te)\cite{A072369,A074775,A080474}. The electronic states near the Fermi surface are dominantly determined by the five $d$-orbitals of Fe in a distorted tetrahedron environment\cite{A032740,A031279,A062630,A074312}, and the prominent nesting feature
of the quasi-two-dimensional Fermi surfaces results in the commensurate SDW ordering for the parent compounds\cite{A033426,A033286} (Fig.~1c). 
Since the superconductivity and commensurate SDW ordering exist in close vicinity\cite{Kamihara2008,A033603,A042053,A054630
}, spin excitations from the SDW order have been proposed as the bosonic ``glue'' producing high $T_C$
superconductivity in these ferrous materials\cite{A033286,A033426,A032740}.

The nesting cylinder-like electron and hole Fermi surfaces, separating by ($\pi$,0) within the Fe square sublattice of the parent compounds, lose the nesting condition when adding electrons or holes to the systems\cite{A043355} since 
such alterations would increase the size of the electron Fermi surface while reducing the size of the hole Fermi surface. 
This expectation is realised in systematic doping\cite{A062528,A063533,A063962,A073950} and pressure studies\cite{A073032}, which show the destruction of the SDW order well before the optimal superconducting state is established. The commensurate SDW order with the same in-plane magnetic propagation vector ($\pi$,0) has also been predicted for $\alpha$-FeTe in a recent band-structure theoretic study\cite{A074312}. However, what we observed directly in neutron scattering experiments is an incommensurate antiferromagnetic order
with the in-plane propagation vector ($\delta\pi, \delta\pi$) along the diagonal direction of the Fe square sublattice (Fig.~1b). Additionally, such an incommensurate magnetic order is robust and survives as short-range correlations in the sample even at the optimized $T_C$.

The binary iron-chalcogen systems exist over the whole composition range, but single phase material in the tetragonal PbO structure exists only in a narrow composition range\cite{mfc_smk}. In the $\alpha$ phase (also called $\beta$ phase in [\cite{mfc_smk}]), iron-chalcogen forms with the same edge-sharing antifluorite layers found in the FeAs high-$T_C$ superconductors. The $\alpha$-FeSe with the nominal composition FeSe$_{0.88}$ was recently reported to superconduct at
$T_C\approx 8$ K\cite{A072369}. The $T_C$ is very sensitive to pressure and increases to 27 K at 1.48 GPa\cite{A074315}. The isovalent series of Fe(Te$_{1-x}$Se$_x$)$_{z}$ in the $\alpha$-phase with nominal $z=0.82$ has been synthesized\cite{A074775}. The $T_C$ is enhanced to 14 K at $x=0.4$ when a substantial fraction of Se is replaced by Te. The end member FeTe$_{0.82}$ is not superconducting. Similar results have also been reported for the nominal $z=1$ series\cite{A080474}.  

The crystal structure of Fe(Te$_{1-x}$Se$_x$)$_{z}$ with excess Fe is shown in Fig.~1a.
The tetragonal structure is well described by the $P4/nmm$ space group (Table 1 in Supplementary Information).
The chalcogen and Fe(1) sites of the PbO structure are fully occupied, and the excess Fe partially occupy the interstitial sites marked as the Fe(2). Thus, instead of being chalcogen deficient, the $\alpha$-iron-chalcogen have an excess of iron, yielding a more appropriate formulas of Fe$_{1+y}$(Te$_{1-x}$Se$_x$). From the refined occupancy, the samples of nominal composition FeTe$_{0.82}$, FeTe$_{0.90}$ and Fe(Te$_{0.6}$Se$_{0.4}$)$_{0.82}$ are actually Fe$_{1.141(2)}$Te,
Fe$_{1.076(2)}$Te and Fe$_{1.080(2)}$Te$_{0.67(2)}$Se$_{0.33(2)}$, respectively. Other parameters determined from Rietveld refinement of neutron powder diffraction spectra are listed in Table 1 in Supplementary Information.
The superconducting sample Fe$_{1.080}$Te$_{0.67}$Se$_{0.33}$ ($T_C\approx 14$ K) remains tetragonal in the superconducting state at 4 K. The parent compounds $\alpha$-FeTe experiences simultaneous structural and magnetic transition, as observed in BaFe$_2$As$_2$\cite{A062776}, though the situation in this case is more complex and discussed below.

The $\alpha$-FeTe exists over a well-defined stoichoimetric range. At low temperatures below a phase transition, two distinct types of transport properties have been observed depending on composition: one is metallic and the other semiconducting (see Supplementary Information). Correspondingly, we find different lattice distortions at the transition for the two types of samples: Fe$_{1.141}$Te is a semiconductor whereas Fe$_{1.076}$Te is metallic at low temperatures despite both possessing the same $P4/nmm$ tetragonal structure at room temperature, as shown in Fig.~2a and c. The Fe$_{1.141}$Te distorts to a $Pmmn$ orthorhombic structure below the transition, as a result of an expansion of one of the in-plane axis and a contraction of the another. This results in the splitting of the ($h0k$) Bragg peaks of the high-temperature structure (see Fig.~2b).
This orthorhombic distortion is different from that observed for the FeAs-based materials in either the ZrCuSiAs or ThCr$_2$Si$_2$ structure which doubles the in-plane unit cell\cite{A040795,A062776}. The distortion of the Fe$_{1.076}$Te is monoclinic, (Fig.~2d). In addition to the expansion and contraction of the in-plane lattice parameters, the $c$-axis rotates towards the $a$-axis. Thus, the monoclinic distortion not only splits the (200) Bragg peaks but also the (112) peak. Refined structure parameters for all compounds are listed in Table 1 in Supplementary Information.

The additional magnetic Bragg reflections in the 8K spectrum of Fe$_{1.141}$Te  (Fig.~2b) cannot be indexed by multiples of the nuclear unit cell. By performing single-crystal neutron diffraction experiments, we determine the magnetic wave-vector as $\mathbf{q}=(\pm \delta, 0, 1/2)$, where $\delta=0.380$. The magnetic wave-vector determines that magnetic moments in each row along the $b$-axis
are aligned parallel to each other (see Fig.~1b). From row to row along the $a$-axis in the same Fe plane, the moments are modulated with the incommensurate propagating vector $\delta \mathbf{a}^*$.
From one plane to the next along the $c$-axis, magnetic moments simply alternate the direction. The magnetic Bragg intensities observed in unpolarized neutron diffraction experiments can be equivalently described by a linear sinusoid model with the magnetic moment along the $b$-axis, a spiral model with the magnetic moment rotating in the $ac$-plane, or a linear combination of the two (see Supplementary Information). Using the polarized neutron scattering, we determined that both the linear and spiral components are present in our sample and that the magnetic structure is defined by:
\begin{equation}
\mathbf{M}(\mathbf{R})={\rm M} [\hat{\mathbf{a}} \cos(\mathbf{q}\cdot \mathbf{R}+\phi_\mathbf{R}) +\hat{\mathbf{c}} \sin(\mathbf{q}\cdot \mathbf{R}+\phi_\mathbf{R})+w\hat{\mathbf{b}} \cos(\mathbf{q}\cdot \mathbf{R}+\phi_\mathbf{R}+\psi)],
\end{equation}
where $\mathbf{R}$ is the position of the Fe in the lattice, ${\rm M}= 0.76(2) \mu_B$ per Fe, $w=2.17(6)$, $\hat{\mathbf{a}}$ is the unit vector of the $a$-axis, $\psi$ is an arbitrary phase between the spiral and the sinusoid components, and $\phi_\mathbf{R}= 0$ for the Fe(1) sites and 112(7)$^o$ for the Fe(2) sites.  The modulus of the magnetic moment oscillates between ${\rm M}$ to $(1+w^2)^{1/2}{\rm M}\approx 2.38 {\rm M}$ in the incommensurate order. This magnetic structure is considerably more complex than the commensurate SDW magnetic order in previous FeAs materials. 
While the antiparallel moment alignment along the $c$-axis is the same, the propagating vector in the plane now is along the diagonal direction of the Fe ``square'' sublattice (Fig.~1b), markedly different from the commensurate SDW which propagates along one edge (Fig.~1c). The projections of the complete magnetic structure of Fe$_{1.141}$Te onto the $ab$ and $ac$ planes are shown in Fig.~S2 in Supplementary Information.

The magnetic and structural transitions occur simultaneously at $T_S\approx 63$ K for Fe$_{1.141}$Te and at $T_S\approx 75$ K for Fe$_{1.076}$Te. In a similar fashion to the first-order transition reported for BaFe$_2$As$_2$\cite{A062776}, thermal hysteresis is present and shown in Fig.~3a and b. The transition is further evidenced by an anomaly in resistivity (Fig.~S1) and magnetic susceptibility at $T_S$\cite{A074775}. In Fig.~3c and d, the temperature dependence of the lattice parameters is shown. At the phase transition, the lattice contracts in the $b$-axis, along which the magnetic moments are parallel to each other, and it expands in the $a$ and $c$-axis, which are the directions of the antiferromagnetic alignment. This is consistent with previously observed magnetostriction pattern in NdFeAsO\cite{A062195} and BaFe$_2$As$_2$\cite{A062776}, and can be understood when the magnetism originates from multiple orbitals\cite{A042252}. Once again, there exists
strong coupling between the lattice and magnetic degrees of freedom. The details of structure evolution as a function of temperature are shown in Fig. S3 in Supplementary Information.

The incommensurate magnetic wave vector can be tuned by varying the sample composition. For Fe$_{1.165(3)}$Te in the same orthorhombic structure, the incommensurability has been greatly affected and measured at $\delta=0.346$, despite no appreciable differences with either the moment or the phase $\phi_\mathbf{R}$ compared to the values for Fe$_{1.141(3)}$Te, see Table 1 in Supplementary Information. For the composition with least excess iron, Fe$_{1.076}$Te, this metallic monoclinic phase locks into a commensurate value $\delta=0.5$. In the inset (a) of Fig.~4, the variation of the $\delta$ as a function of the excess Fe
$y$ is shown.

Having determined the incommensurate magnetic structure in the parent compound $\alpha$-FeTe, a natural question is whether the new magnetic order is relevant
to the superconducting samples. To answer the question, we chose the nominal $x=0.4$ composition of the superconductor series Fe(Te$_{1-x}$Se$_x$)$_z$ as it possesses the optimal $T_C \approx 14$ K\cite{A074775}. The refined composition is Fe$_{1.080(2)}$Te$_{0.67(2)}$Se$_{0.33(2)}$. There is neither long-range magnetic order nor structural transition in this sample, though we observed pronounced short-range magnetic correlations
at the incommensurate wave-vector $(0.438,0,1/2)$ (Fig.~4). The half-width-at-the-half-maximum of 0.25 $\AA^{-1}$ indicates a magnetic correlation length of 4.0 $\AA$, which equates approximately to two nearest-neighbour Fe spacings.
The concave shape of the peak intensity as a function of temperature in the inset (b) indicates the expected diffusive nature for
the short-range magnetic correlations, in contrast to the convex functional form for long-range magnetic order (Fig.~3a and b). This is very different from the case of the commensurate SDW which is completely suppressed
in the optimal $T_C$ samples\cite{A040795,A062528,A063533,A073950}.

To summarize, the $\alpha$-iron-chalcogen in the PbO structure shares the same fundamental structural building block and nesting electronic band structure\cite{A074312} as in previously reported for the FeAs-based superconductor systems. Though the same commensurate SDW magnetic order has been predicted, we show the presence of a fundamentally different incommensurate antiferromagnetic order which propagates with wave vector ($\delta\pi,\delta\pi$) along the diagonal direction. This long ranged magnetic order is easily tunable with composition and locks into a commensurate value ($\pi/2,\pi/2$) in the metallic phase. This order cannot be the result of nesting Fermi surface as this is along the ($\pi,0$) direction and delicately depends on electronic band filling for its existence. The incommensurate magnetic order survives as short-range magnetic
correlations even in the superconducting state. These results demonstrate that there are other magnetic modes other than the vulnerable ($\pi,0$) SDW mode to mediate the Cooper pairs in the ferrous superconductors\cite{A043355}.

\begin{methods}
We synthesized polycrystalline samples, weighing 15-16 g for each of the Fe$_{1.141}$Te, Fe$_{1.076}$Te and Fe$_{1.080}$(Te$_{0.67}$Se$_{0.33}$) and 4 g for Fe$_{1.165}$Te, using the solid state reaction method. The single-crystal sample of Fe$_{1.14}$Te was grown using a flux
method and shows the similar behavior in the resistivity and magnetization measurements to those obtained for the polycrystalline samples.

Temperature-dependent magnetic Bragg diffraction on polycrystalline samples was measured using the cold neutron triple-axis spectrometer SPINS, and using the single-crystal sample at thermal neutron triple-axis spectrometer BT9 at the NIST Center for Neutron Research (NCNR). The high resolution powder diffraction spectra for refinements were measured using BT1, and
the short-range magnetic correlations of Fe$_{1.080}$(Te$_{0.67}$Se$_{0.33}$) were measured using DCS at NCNR. The spin-dependent partial cross-sections of the single crystal sample were measured using the Asterix spectrometer at the Lujan Center of Los Alamos National Laboratory.  The sample temperature was controlled by a pumped He cryostat at SPINS, DCS and BT9, and by a Displex refrigerator at BT1 and Asterix.

%
%
%
\end{methods}



\bibliography{/home/bao/kept/tex/bib4/FeAs,/home/bao/kept/tex/bib4/frus,/home/bao/kept/tex/bib4/ruth}

\begin{thebibliography}{10}
\expandafter\ifx\csname url\endcsname\relax
  \def\url#1{\texttt{#1}}\fi
\expandafter\ifx\csname urlprefix\endcsname\relax\def\urlprefix{URL }\fi
\providecommand{\bibinfo}[2]{#2}
\providecommand{\eprint}[2][]{\url{#2}}

\bibitem{Kamihara2008}
\bibinfo{author}{Kamihara, Y.}, \bibinfo{author}{Watanabe, T.},
  \bibinfo{author}{Hirano, M.} \& \bibinfo{author}{Hosono, H.}
\newblock \bibinfo{title}{Iron-based layered superconductor
  La[O$_{1-x}$F$_x$]FeAs ($x$=0.05 - 0.12) with $T_c$=26K}.
\newblock \emph{\bibinfo{journal}{J.\ Am.\ Chem.\ Soc.}}
  \textbf{\bibinfo{volume}{130}}, \bibinfo{pages}{3296} (\bibinfo{year}{2008}).

\bibitem{A033603}
\bibinfo{author}{Chen, X.~H.} \emph{et~al.}
\newblock \bibinfo{title}{Superconductivity at 43 K in SmFeAsO$_{1-x}$F$_x$}.
\newblock \emph{\bibinfo{journal}{Nature}} \textbf{\bibinfo{volume}{453}},
  \bibinfo{pages}{761} (\bibinfo{year}{2008}).

\bibitem{A042053}
\bibinfo{author}{Ren, Z.~A.} \emph{et~al.}
\newblock \bibinfo{title}{Superconductivity at 55 K in iron-based F-doped
  layered quaternary compound Sm[O$_{1-x}$F$_x$]FeAs}.
\newblock \emph{\bibinfo{journal}{Chinese Phys. Lett.}}
  \textbf{\bibinfo{volume}{25}}, \bibinfo{pages}{2215} (\bibinfo{year}{2008}).

\bibitem{A054630}
\bibinfo{author}{Rotter, M.}, \bibinfo{author}{Tegel, M.} \&
  \bibinfo{author}{Johrendt, D.}
\newblock \bibinfo{title}{Superconductivity at 38 K in the iron arsenide
  (Ba$_{1-x}$K$_x$)Fe$_2$As$_2$}.
\newblock \emph{\bibinfo{journal}{arXiv:0805.4630}}  (\bibinfo{year}{2008}).

\bibitem{A040795}
\bibinfo{author}{de~la Cruz, C.} \emph{et~al.}
\newblock \bibinfo{title}{Magnetic order close to superconductivity in the
  iron-based layered La(O$_{1-x}$F$_x$)FeAs systems}.
\newblock \emph{\bibinfo{journal}{Nature}} \textbf{\bibinfo{volume}{453}},
  \bibinfo{pages}{899} (\bibinfo{year}{2008}).

\bibitem{A062776}
\bibinfo{author}{Huang, Q.} \emph{et~al.}
\newblock \bibinfo{title}{Magnetic order in BaFe$_2$As$_2$, the parent compound
  of the FeAs based superconductors in a new structural family}.
\newblock \emph{\bibinfo{journal}{arXiv:0806.2776}}  (\bibinfo{year}{2008}).

\bibitem{A033286}
\bibinfo{author}{Ma, F.} \& \bibinfo{author}{Lu, Z.~Y.}
\newblock \bibinfo{title}{Iron-based layered superconductor
  LaFeAsO$_{1-x}$F$_x$: an antiferromagnetic semimetal}.
\newblock \emph{\bibinfo{journal}{Phys. Rev. B}} \textbf{\bibinfo{volume}{78}},
  \bibinfo{pages}{033111} (\bibinfo{year}{2008}).

\bibitem{A033426}
\bibinfo{author}{Dong, J.} \emph{et~al.}
\newblock \bibinfo{title}{Competing orders and spin-density-wave instability in
  La(O$_{1-x}$F$_x$)FeAs}.
\newblock \emph{\bibinfo{journal}{Europhys.\ Lett.}}
  \textbf{\bibinfo{volume}{83}}, \bibinfo{pages}{27006} (\bibinfo{year}{2008}).

\bibitem{A043355}
\bibinfo{author}{Yin, Z.~P.} \emph{et~al.}
\newblock \bibinfo{title}{Electron-hole symmetry and magnetic coupling in
  antiferromagnetic LaOFeAs}.
\newblock \emph{\bibinfo{journal}{Phys. Rev. Lett.}}
  \textbf{\bibinfo{volume}{101}}, \bibinfo{pages}{047001}
  (\bibinfo{year}{2008}).

\bibitem{A072369}
\bibinfo{author}{Hsu, F.-C.} \emph{et~al.}
\newblock \bibinfo{title}{Superconductivity in the PbO-type structure
  $\alpha$-FeSe}.
\newblock \emph{\bibinfo{journal}{arXiv:0807.2369}}  (\bibinfo{year}{2008}).

\bibitem{A074775}
\bibinfo{author}{Fang, M.} \emph{et~al.}
\newblock \bibinfo{title}{Superconductivity close to magnetic instability in
  Fe(Se$_{1-x}$Te$_x$)$_{0.82}$}.
\newblock \emph{\bibinfo{journal}{arXiv:0807.4775}}  (\bibinfo{year}{2008}).

\bibitem{A080474}
\bibinfo{author}{Yeh, K.-W.} \emph{et~al.}
\newblock \bibinfo{title}{Tellurium substitution effect on superconductivity of
  the $\alpha$-phase Iron Selenide}.
\newblock \emph{\bibinfo{journal}{arXiv:0808.0474}}  (\bibinfo{year}{2008}).

\bibitem{A074312}
\bibinfo{author}{Subedi, A.}, \bibinfo{author}{Zhang, L.},
  \bibinfo{author}{Singh, D.} \& \bibinfo{author}{Du, M.}
\newblock \bibinfo{title}{Density functional study of FeS, FeSe and FeTe:
  Electronic structure, magnetism, phonons and superconductivity}.
\newblock \emph{\bibinfo{journal}{arXiv:0807.4312}}  (\bibinfo{year}{2008}).

\bibitem{mag_sc}
\bibinfo{author}{Abrikosov, A.~A.} \& \bibinfo{author}{Gorkov, L.~P.}
\newblock \emph{\bibinfo{journal}{Zh. Eksp. Teor. Fiz.}}
  \textbf{\bibinfo{volume}{39}}, \bibinfo{pages}{1781} (\bibinfo{year}{1960}).

\bibitem{A061209}
\bibinfo{author}{Chen, G.~F.} \emph{et~al.}
\newblock \bibinfo{title}{Superconductivity in hole-doped
  (Sr$_{1-x}$K$_x$)Fe$_2$As$_2$}.
\newblock \emph{\bibinfo{journal}{arXiv:0806.1209}}  (\bibinfo{year}{2008}).

\bibitem{A061301}
\bibinfo{author}{Sasmal, K.} \emph{et~al.}
\newblock \bibinfo{title}{Superconductivity up to 37 K in
  (A$_{1-x}$Sr$_x$)Fe2As2 with A = K and Cs}.
\newblock \emph{\bibinfo{journal}{arXiv:0806.1301}}  (\bibinfo{year}{2008}).

\bibitem{A064279}
\bibinfo{author}{Wu, G.} \emph{et~al.}
\newblock \bibinfo{title}{Different resistivity response to spin density wave
  and superconductivity at 20 K in Ca$_{1-x}$Na$_x$Fe$_2$As$_2$}.
\newblock \emph{\bibinfo{journal}{arXiv:0806.4279}}  (\bibinfo{year}{2008}).

\bibitem{A032740}
\bibinfo{author}{Mazin, I.~I.}, \bibinfo{author}{Singh, D.~J.},
  \bibinfo{author}{Johannes, M.~D.} \& \bibinfo{author}{Du, M.~H.}
\newblock \bibinfo{title}{Unconventional sign-reversing superconductivity in
  LaFeAsO$_{1-x}$F$_x$}.
\newblock \emph{\bibinfo{journal}{Phys. Rev. Lett.}}
  \textbf{\bibinfo{volume}{101}}, \bibinfo{pages}{057003}
  (\bibinfo{year}{2008}).

\bibitem{A031279}
\bibinfo{author}{Haule, K.}, \bibinfo{author}{Shim, J.~H.} \&
  \bibinfo{author}{Kotliar, G.}
\newblock \bibinfo{title}{Correlated electronic structure of
  LaFeAsO$_{1-x}$F$_x$}.
\newblock \emph{\bibinfo{journal}{Phys. Rev. Lett.}}
  \textbf{\bibinfo{volume}{100}}, \bibinfo{pages}{226402}
  (\bibinfo{year}{2008}).

\bibitem{A062630}
\bibinfo{author}{Nekrasov, I.~A.}, \bibinfo{author}{Pchelkina, Z.~V.} \&
  \bibinfo{author}{Sadovskii, M.~V.}
\newblock \bibinfo{title}{Electronic structure of prototype AFe$_2$As$_2$ and
  ReOFeAs high-temperature superconductors: a comparison}.
\newblock \emph{\bibinfo{journal}{arXiv:0806.2630}}  (\bibinfo{year}{2008}).

\bibitem{A062528}
\bibinfo{author}{Zhao, J.} \emph{et~al.}
\newblock \bibinfo{title}{Structural and magnetic phase diagram of
  CeFeAs$_{1-x}$F$_x$ and its relationship to high-temperature
  superconductivity}.
\newblock \emph{\bibinfo{journal}{arXiv:0806.2528}}  (\bibinfo{year}{2008}).

\bibitem{A063533}
\bibinfo{author}{Luetkens, H.} \emph{et~al.}
\newblock \bibinfo{title}{Electronic phase diagram of the LaO$_{1-x}$F$_x$FeAs
  superconductor}.
\newblock \emph{\bibinfo{journal}{arXiv:0806.3533}}  (\bibinfo{year}{2008}).

\bibitem{A063962}
\bibinfo{author}{Margadonna, S.} \emph{et~al.}
\newblock \bibinfo{title}{Crystal structure and phase transitions across the
  metal-superconductor boundary in the SmFeAsO$_{1-x}$F$_x$ ($0 \le x \le
  0.20$) family}.
\newblock \emph{\bibinfo{journal}{arXiv:0806.3962}}  (\bibinfo{year}{2008}).

\bibitem{A073950}
\bibinfo{author}{Chen, H.} \emph{et~al.}
\newblock \bibinfo{title}{Coexistence of the spin-density-wave and
  superconductivity in the Ba$_{1-x}$K$_x$Fe$_2$As$_2$}.
\newblock \emph{\bibinfo{journal}{arXiv:0807.3950}}  (\bibinfo{year}{2008}).

\bibitem{A073032}
\bibinfo{author}{Kreyssig, A.} \emph{et~al.}
\newblock \bibinfo{title}{Squeezing the magnetism out of superconducting
  CaFe$_2$As$_2$}.
\newblock \emph{\bibinfo{journal}{arXiv:0807.3032}}  (\bibinfo{year}{2008}).

\bibitem{mfc_smk}
\bibinfo{author}{Schuster, W.}, \bibinfo{author}{Mikler, H.} \&
  \bibinfo{author}{Komarek, K.}
\newblock \bibinfo{title}{Transition metal-chalcogen systems, VII: the
  iron-selenium phase diagram}.
\newblock \emph{\bibinfo{journal}{Monatshefte f{\"{u}}r Chem.}}
  \textbf{\bibinfo{volume}{110}}, \bibinfo{pages}{1153} (\bibinfo{year}{1979}).

\bibitem{A074315}
\bibinfo{author}{Mizuguchi, Y.}, \bibinfo{author}{Tomioka, F.},
  \bibinfo{author}{Tsuda, S.}, \bibinfo{author}{Yamaguchi, T.} \&
  \bibinfo{author}{Takano, Y.}
\newblock \bibinfo{title}{Superconductivity at 27 K in tetragonal FeSe under
  high pressure}.
\newblock \emph{\bibinfo{journal}{arXiv:0807.4315}}  (\bibinfo{year}{2008}).

\bibitem{A062195}
\bibinfo{author}{Qiu, Y.} \emph{et~al.}
\newblock \bibinfo{title}{Structure and magnetic order in the
  NdFeAsO$_{1-x}$F$_x$ superconductor system}.
\newblock \emph{\bibinfo{journal}{arXiv:0806.2195}}  (\bibinfo{year}{2008}).

\bibitem{A042252}
\bibinfo{author}{Yildirim, T.}
\newblock \bibinfo{title}{Origin of the 150 K Anomaly in LaOFeAs; Competing
  Antiferromagnetic Superexchange Interactions, Frustration, and Structural
  Phase Transition}.
\newblock \emph{\bibinfo{journal}{Phys. Rev. Lett.}}
  \textbf{\bibinfo{volume}{101}}, \bibinfo{pages}{057010}
  (\bibinfo{year}{2008}).

\end{thebibliography}


\begin{addendum}
\item Work at LANL is supported by U.S.\ DOE's Office of Basic Energy Science; at Tulane by the NSF under grant DMR-0645305, the DOE under DE-FG02-07ER46358 and the Research Corporation; at ZU by National Basic Research Program of China (No.2006CB01003, 2009CB929104) and the PCSIRT of the Ministry of Education of China (IRT0754); at UNO is by DARPA through Grant No. HR0011-07-1-0031.
The SPINS and DCS at NIST is partially supported by NSF under Agreement No. DMR-0454672. LANL is operated by LANS LLC under DOE Contract DE-AC52-06NA25396. 
 \item[Author information] The authors declare that they have no
competing financial interests.
 Correspondence and requests for materials
should be addressed to W.B. (wbao@lanl.gov).
\end{addendum}


\newpage

\begin{figure}
\vskip -2cm
\begin{center}
\includegraphics[scale=.8,angle=0,clip=true]{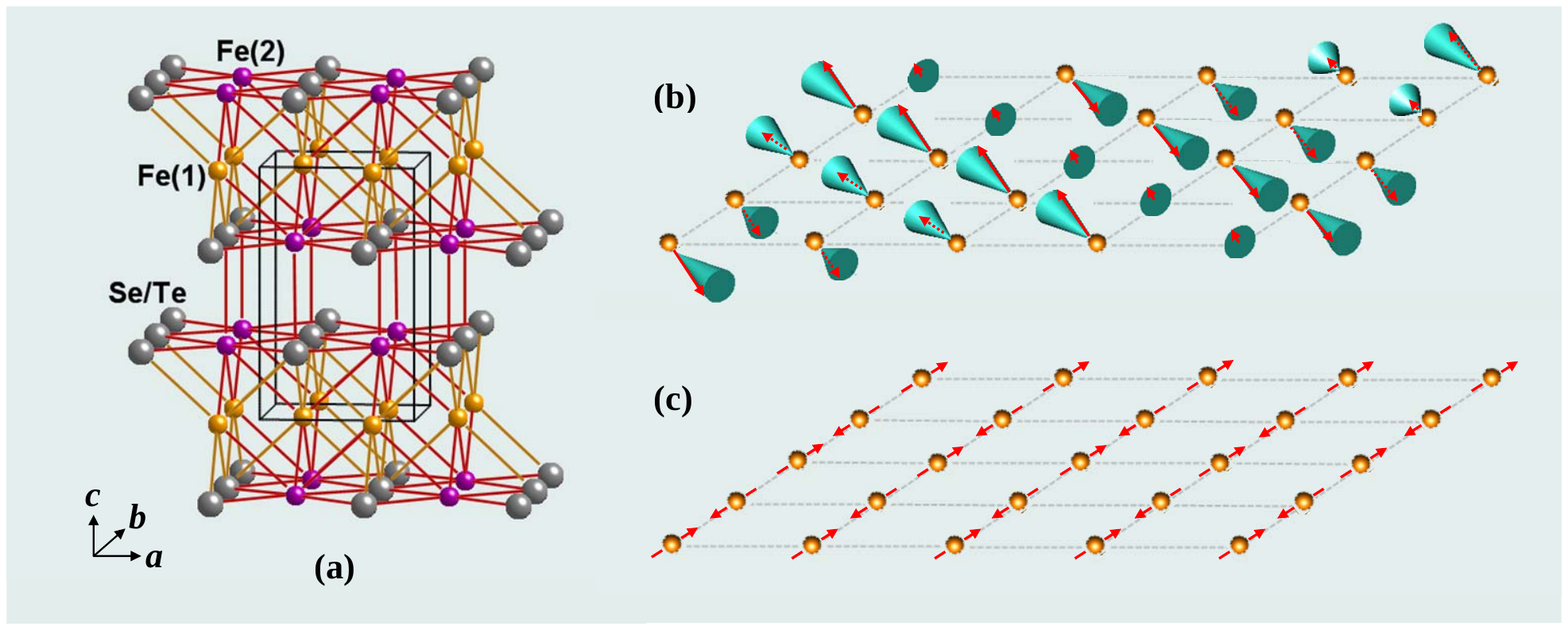}
\end{center}
\vskip -7.5cm
\caption{{\bf Crystal and magnetic structures of $\alpha$-iron-chalcogen.} {\bf a}, The Fe(1) and Te/Se are fully occupied sites of the tetragonal PbO structure. The excessive Fe of Fe$_{1+y}$(Te$_{1-x}$Se$_x$) is located on the partially occupied Fe(2) site. Magnetic structures of {\bf b}, $\alpha$-FeTe and {\bf c}, BaFe$_2$As$_2$ are shown in a Fe layer for comparison. The dashed lines denotes the Fe ``square'' sublattice. Each row of magnetic moments along the diagonal direction (the $b$-axis of the crystal structure) are identical in {\bf b}. They modulate incommensurately with the lattice along the other diagonal (the $a$-axis), with both a linear sinusoid component pointing in the $b$-direction and a spiral component in the $ac$-plane. In contrast, in the commensurate antiferromagnetic order of BaFe$_2$As$_2$, the row of identical magnetic moments line along one edge of the Fe square lattice (the $a$-axis of the orthorhombic unit cell) and alternate along the other edge (the $b$-axis)\cite{A062776}. }
\label{fig1}
\end{figure}


\begin{figure}
\begin{center}
\includegraphics[scale=.25,angle=0,clip=true]{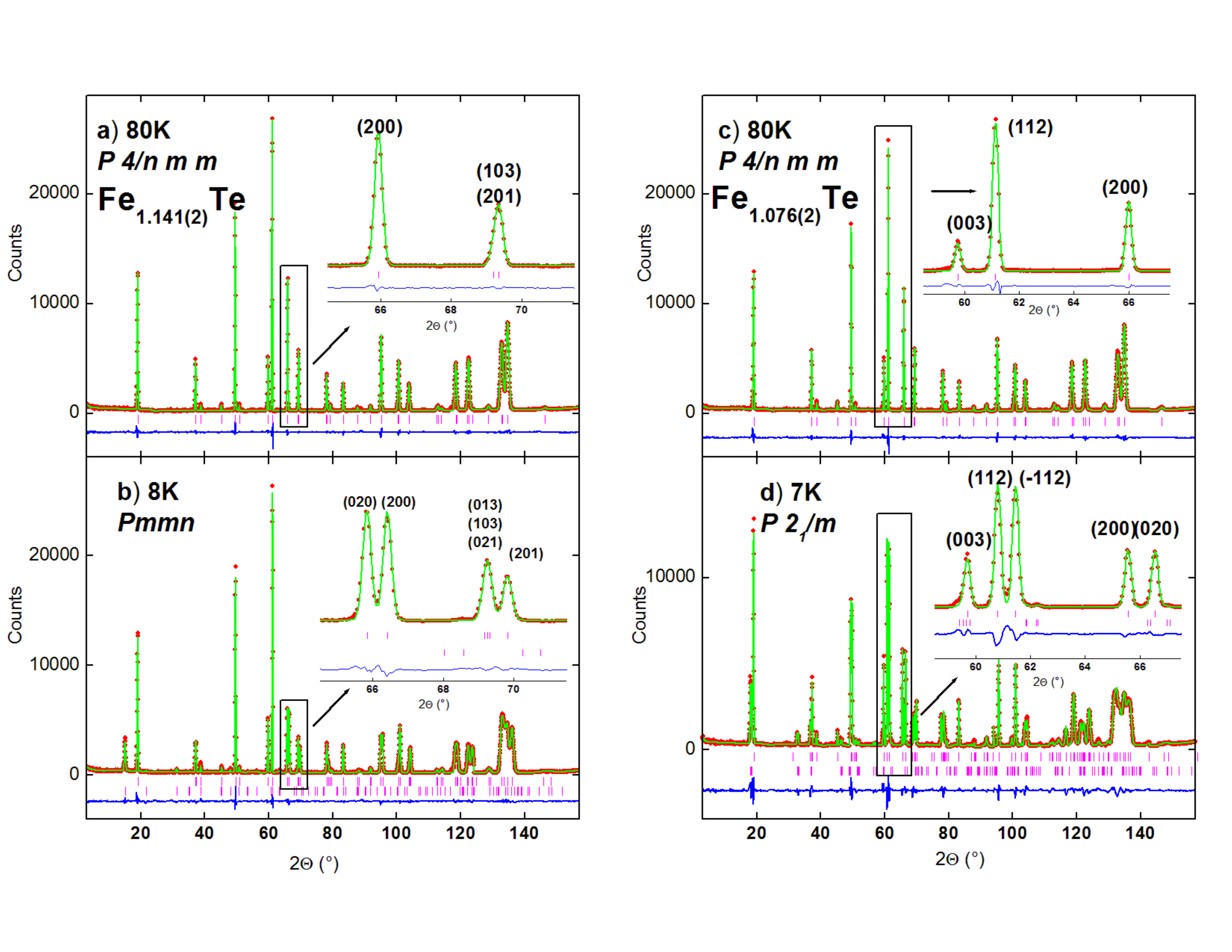}
\end{center}
\vskip -9cm
\caption{{\bf Neutron powder diffraction spectra of Fe$_{1.141}$Te and Fe$_{1.076}$Te.}  Measured with neutrons of wavelength 2.0785 $\AA$. {\bf a},{\bf c}: The high temperature crystal structure is refined with the tetragonal $P4/nmm$ space group; the low temperature one with {\bf b} the orthorhombic $Pmmn$ and {\bf d} the monoclinic $P2_1/m$ space group, respectively. The second row of vertical lines in {\bf b} and {\bf d} mark the magnetic peak positions. The insets highlight splitting of the Bragg peaks by the change of the crystal symmetry. }
\label{fig2}
\end{figure}

\newpage

\begin{figure}
\begin{center}
\includegraphics[width=15cm,angle=0,clip=true]{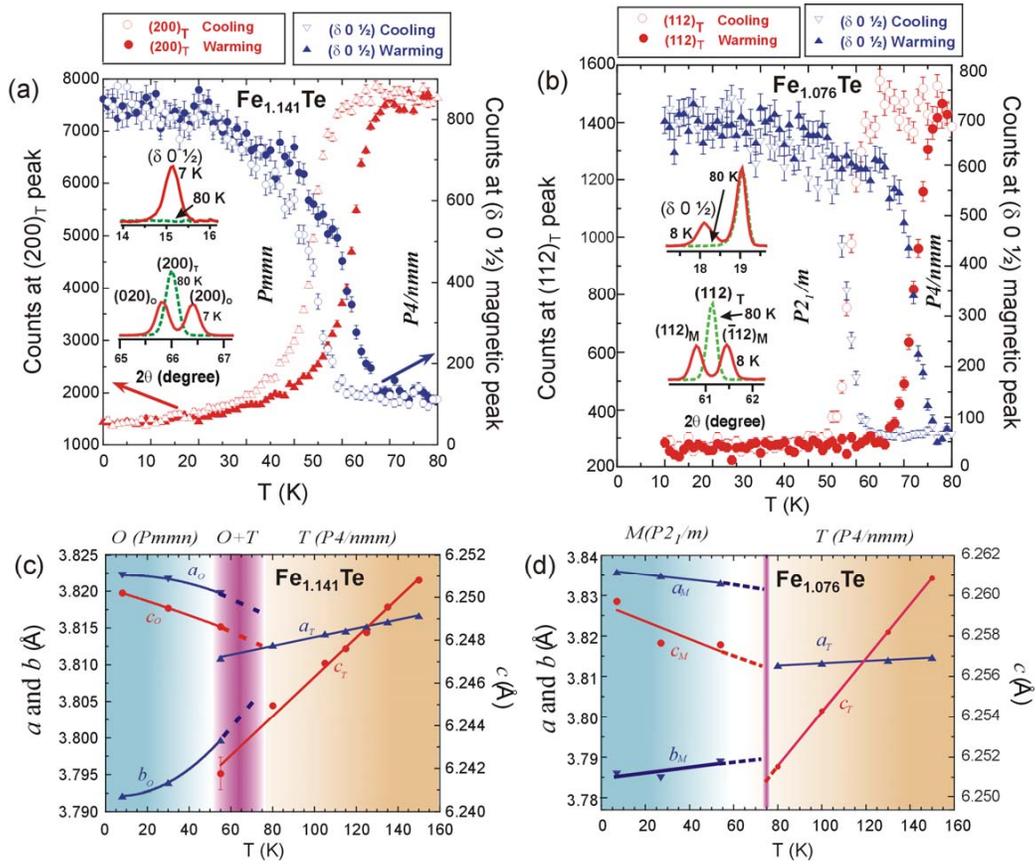}
\end{center}
\vskip -8cm
\caption{{\bf Simultaneous magnetic and structure transition}. {\bf a,b}, The splitting of the structural Bragg peak as a function of temperature was monitored at the peak position of the high-temperature tetragonal phase (red symbols). The intensity of the magnetic Bragg peak at $(\delta,0,1/2)$ (blue symbols) shows the similar thermal hysteresis as the structure peak in the first-order transition.
{\bf c,d}, The lattice parameters as a function of temperature. The $a$ and $c$ axes expand while the $b$ axis contracts.
For Fe$_{1.141}$Te warming up to 55 K, there exist both the orthorhombic (85\%) and tetragonal (15\%) structural components.}
\label{fig5}
\end{figure}

\newpage

\begin{figure}
\begin{center}
\includegraphics[width=14cm,angle=90,clip=true]{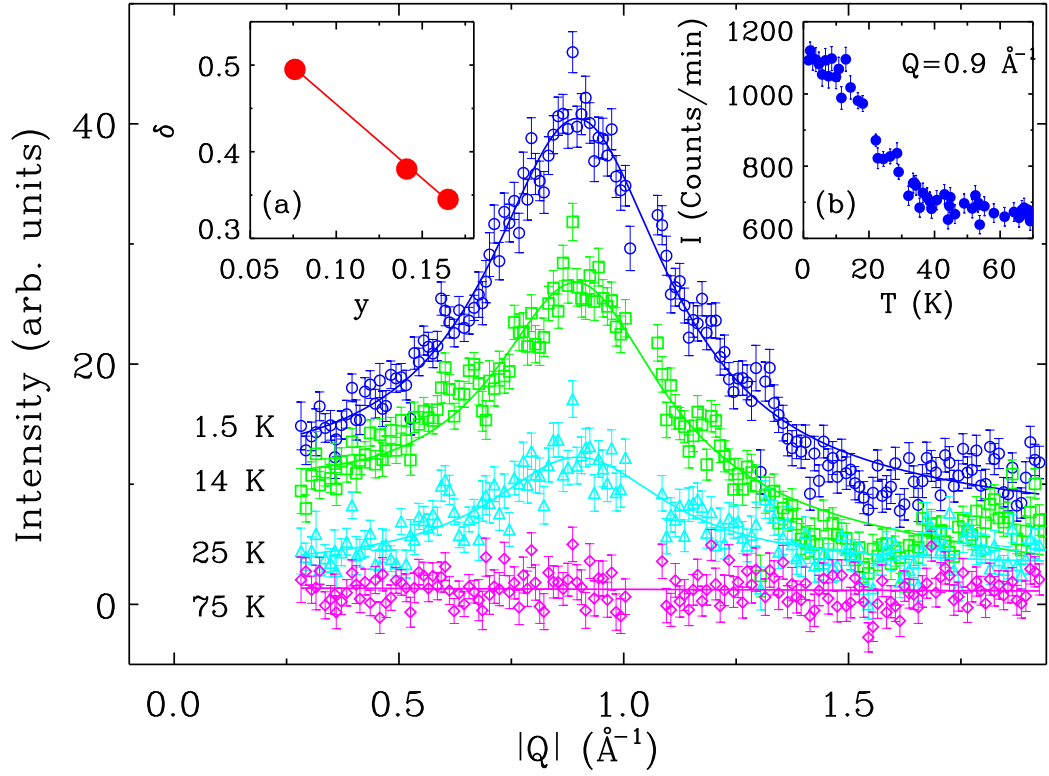}
\end{center}
\vskip -8cm
\caption{{\bf Short-range magnetic order in superconducting Fe$_{1.080(2)}$Te$_{0.67(2)}$Se$_{0.33(2)}$}. It peaks at $q=0.895(3) \AA^{-1} = |(0.438,0,1/2)|$. The width of the peak indicates a magnetic correlation length of 4.0(1) $\AA$. The peak intensity as a function of temperature, inset (b), shows the gradual development of the short-range magnetic order below $\sim$50 K. Inset (a) shows the incommensurability $\delta$ as a function of $y$ for the parent compound Fe$_{1+y}$Te. }
\label{fig6}
\end{figure}

\newpage

\section*{Supplementary Information}

\setcounter{equation}{0}
\renewcommand{\theequation}{S\arabic{equation}}
\setcounter{figure}{0}
\renewcommand{\thefigure}{Figure S\arabic{figure}}

\subsection{\sf Resistivity of the $\alpha$-FeTe.}

We successfully synthesize $\alpha$-FeTe$_x$ in the PbO structure for nominal $x$ from 0.78 to 1.20 using solid state reaction. The resistivity of these parent compounds was measured using the standard four-probes method, and is shown in Fig.~S1.
Two distinct types of resistivity curves have been observed:
the $y=0.78$ and 0.82 samples remains semiconductors below the phase-transition at $T_S\sim$63 K, while the $y \ge 0.90$ samples experience a first-order metal-insulator transition at $T_S\sim$75 K. Therefore, there are two different low-temperature phases for the $\alpha$-FeTe, although the materials share the same tetragonal PbO structure above the phase transition.

\begin{figure}
\vskip -15cm
\includegraphics[scale=.9,angle=0,clip=true]{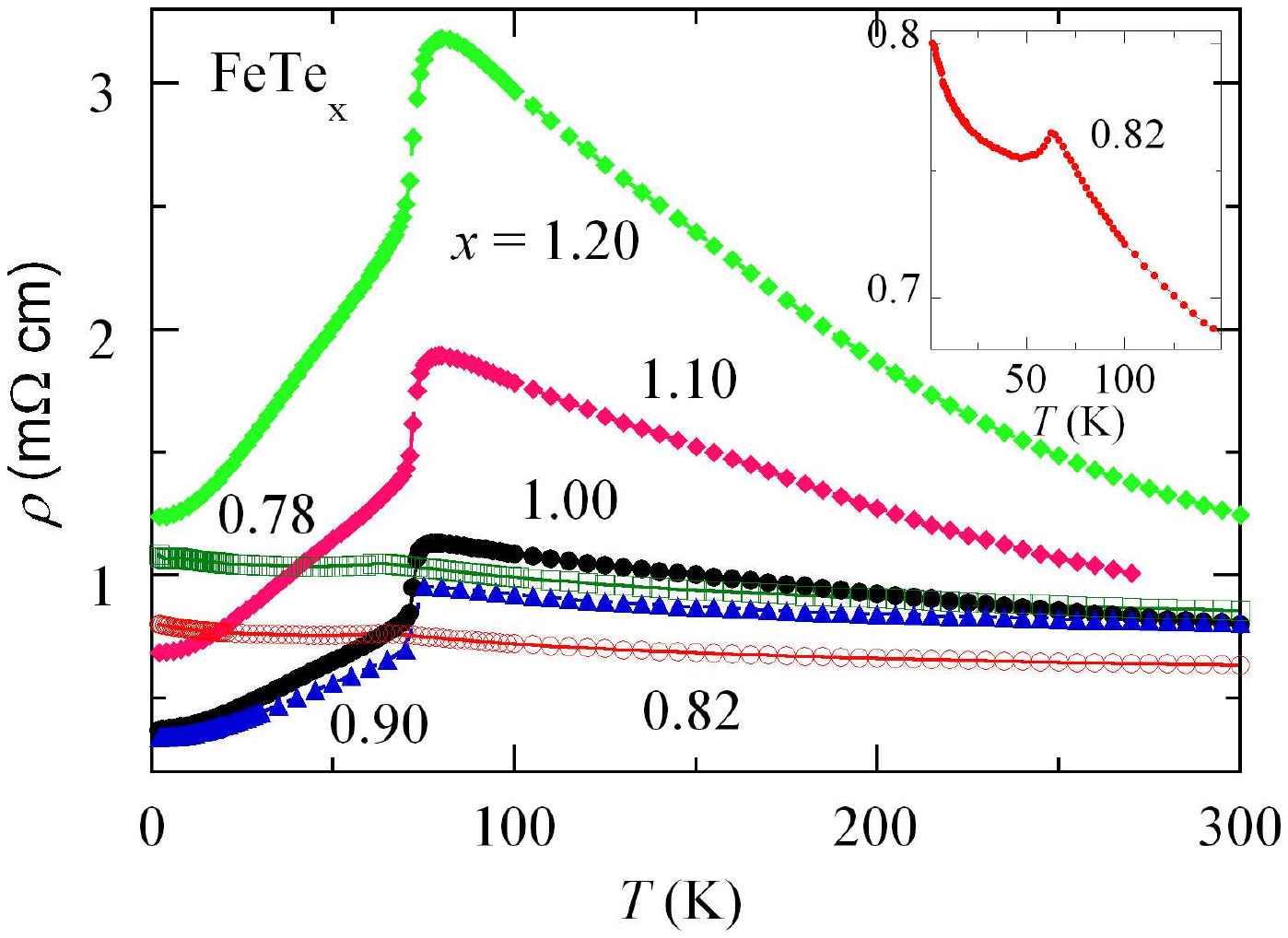}
\vskip -1.2cm
\caption{ Resistivity of nominal FeTe$_x$ as a function of temperature. The inset highlights resistivity anomaly at $T_S$ for the $x=0.82$ sample.}
\end{figure}

\newpage

\subsection{\sf Determination of the incommensurate magnetic structure.}

Magnetic Bragg peaks in the powder neutron diffraction spectra of Fe$_{1.141}$Te cannot be indexed by simple doubling of the crystalline unit cell. Several schemes with incommensurate magnetic vector propagating in different reciprocal directions are possible. To resolve the issue, a single-crystal sample of $\sim$1/8 g was used at the triple-axis spectrometer BT9, using $E_i=E_f=14.7$ meV, to search for magnetic Bragg peaks in the ($hhl$) and ($h0l$) scattering plane. Incommensurate magnetic peaks ($\pm \delta,0,n/2$), n=1,3,5,7,9; ($2-\delta,0,n/2$), n=1,3,5; ($2+\delta,0,n/2$), n=1,3; ($1\pm \delta,0,n/2$), n=1,3; $\delta=0.38$ were observed at 1.6 K. Therefore the magnetic wave-vector is $\mathbf{q}=(\pm \delta, 0, 1/2)$.

The integrated intensity show little angular dependence in the ($h0l$) plane,
suggesting a linear sinusoid magnetic order with the moment perpendicular to the plane, i.e.,
\begin{equation}
\mathbf{M}(\mathbf{R})={\rm M}_l \hat{\mathbf{b}} \cos(\mathbf{q}\cdot \mathbf{R}+\phi_\mathbf{R}),
\end{equation}
its equivalent spiral magnetic order with the moment rotating in the plane,
\begin{equation}
\mathbf{M}(\mathbf{R})={\rm M}_s [\hat{\mathbf{a}} \cos(\mathbf{q}\cdot \mathbf{R}+\phi_\mathbf{R}) +\hat{\mathbf{c}} \sin(\mathbf{q}\cdot \mathbf{R}+\phi_\mathbf{R})],
\end{equation}
or their linear combinations. Note that there is a relation between the neutron scattering cross-sections of the model S1 and S2
\begin{equation}
2\sigma_{S1}({\bf Q})/<{\rm M}_l>^2 =\sigma_{S2}({\bf Q})/<{\rm M}_s>^2,
\end{equation}
therefore, the two models cannot be distinguished in an unpolarized neutron experiment. On the other hand, $\sigma_{S1}=\sigma^{bb}$ and $\sigma_{S2}$ is the $ac$-plane partial cross-section. Thus, they can be readily separated in a polarized neutron measurement of, e.g., the $(\delta, 0, 1/2)$ magnetic Bragg peak.

The model S1 or S2 is confirmed by independent Rietveld refinement of the 8 K powder diffraction spectrum using the Fullprof program. Furthermore, the phase $\phi_\mathbf{R}$ are determined in the refinement. Allowing different moment sizes for Fe(1) and Fe(2) does not significantly improve the fitting. Therefore, the same moment size for each iron was used in the final combined structural and magnetic refinement of the low temperature powder spectra and two examples are shown in Fig.~2b and d. Results using the model S1 are listed in the Table 1.

To determine the contributions of $\sigma_{S1}$ and $\sigma_{S2}$ to magnetic neutron scattering, polarized neutron scattering experiments were performed at Asterix. A single-crystal sample was aligned in the ($h0l$) scattering plane and the neutron spin is controlled by a guide field to align either perpendicular to the ($h0l$) plane (the VF configuration) or parallel to the momentum transfer (the HF configuration). All four channels (++, +-, -+, -\,-) in both configurations were measured for the (001) and $(\delta, 0, 1/2)$ Bragg peaks. The magnetic nature of the $(\delta, 0, 1/2)$ is proved by the spin-flip scattering in the HF configuration. The flipping ratio of the instrumental setup in the VF configuration is 10.3 measured at structural (001) peak. The normalized intensity of $(\delta, 0, 1/2)$ is 8.24(28) in the non-spin-flip (NSF) channels, and 4.13(20) in the spin-flip (SF) channels. After correcting for the finite flipping ratio 10.3, the $\sigma_{S1}=I_{NSF}=7.91(27)$ and $\sigma_{S2}=I_{SF}=3.37(16)$. 

Using Eq.~(S3), 
\begin{equation}
 {\rm M}_l/{\rm M}_s=(2\sigma_{S1}/\sigma_{S2})^{1/2}=2.17(6).
\end{equation}
This is the constant $w\equiv {\rm M}_l/{\rm M}_s$ in Eq.~(1). The ${\rm M}$ in Eq.~(1) relates to the magnetic moment $M$ listed in Table 1 using only the S1 component by
\begin{equation}
{\rm M}=M/(2+w^2)^{1/2},
\end{equation}
since $M^2\equiv {\rm M}_l^2+2{\rm M}_s^2=(w^2+2){\rm M}_s^2$.
In other words, the final result for the incommensurate magnetic structure in Eq.~(1) is a summation of model S1 and S2 with the weight $w:1$.
The projection onto the $ab$ plane is the S1 component, and onto the $ac$ plane the S2 component, see Fig.~S2.

\begin{figure}
\includegraphics[width=15cm,angle=0,clip=true]{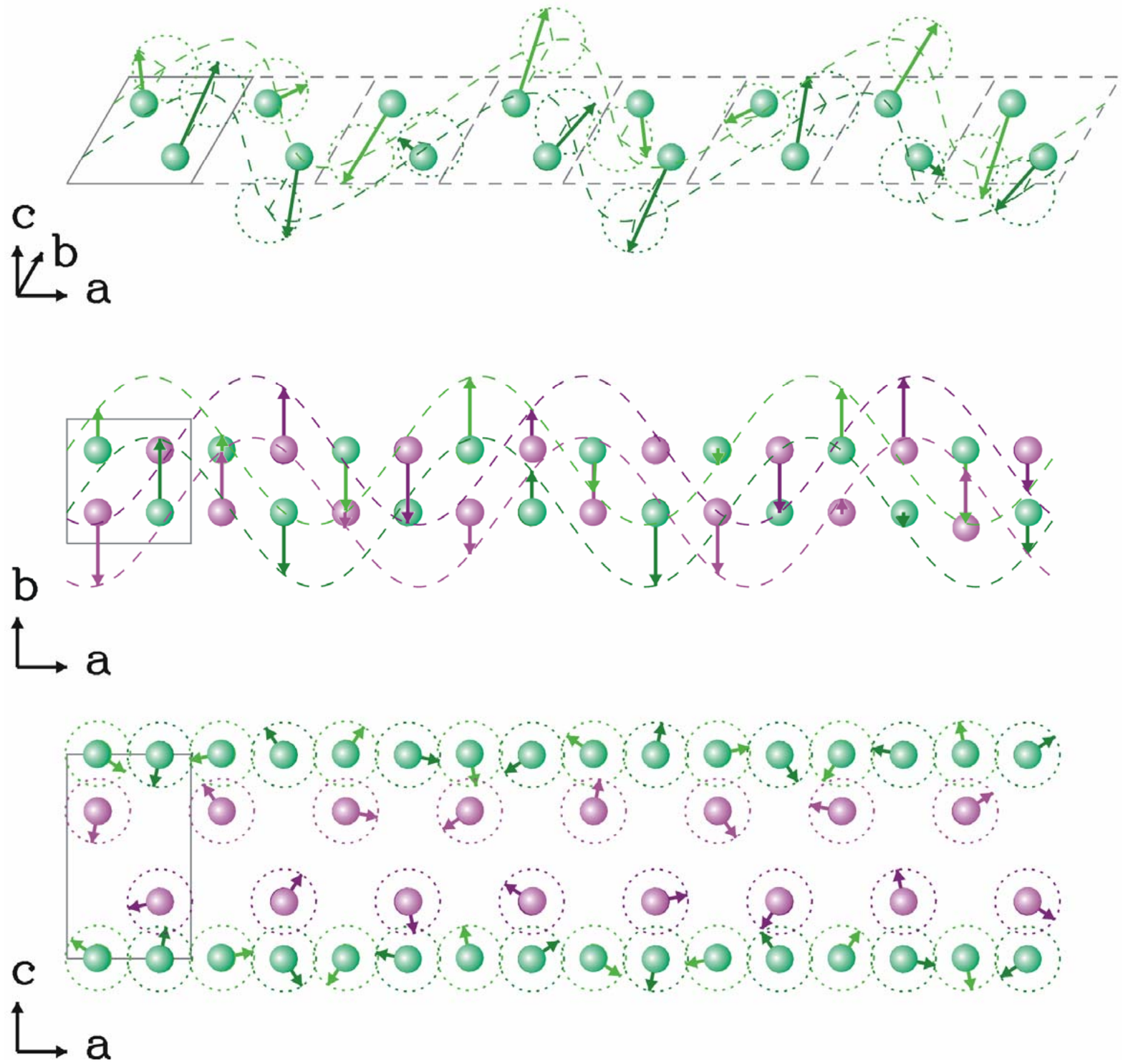}
\vskip -5cm
\caption{{\bf Magnetic structure of $\alpha$-FeTe.}
Top, A three-dimensional view of a single FeTe layer. In the projections,
the excess Fe (purple) are also shown. The structural unit cell is marked by the Gray line.
}
\label{fig3}
\end{figure}

\newpage

\subsection{\sf Refined structure parameters of Fe$_{1+y}$(Te$_{1-x}$Se$_x$).}

The crystal structure is refined using the GSAS program. The combined magnetic and crystal structure refinements were performed using the Fullprof suite of programs.
The structure parameters at selected temperatures are listed in Table~1. 

Detailed temperature evolution of the structure was investigated for the two
representative parent compounds Fe$_{1.141(2)}$Te and
Fe$_{1.076(2)}$Te.
The lattice parameters first decrease linearly as the temperature 
decreases, see Fig 3(c)(d). At the simultaneous magnetic and structural phase-transition, 
the $a$ and $c$ axes expand while the $b$ axis contracts. The lattice distortion is larger for
Fe$_{1.076(2)}$Te, which also experiences a metal-insulator transition at the same temperature.
The transition in Fe$_{1.076(2)}$Te is obviously first-order. The transition in Fe$_{1.141(2)}$Te
is less sharp, and there exist a coexistence range indicated by magenta where
both the orthorhombic(85\%) and tetragonal (15\%) structural components are observed.

The nearest neighbour spacing between Fe(1) ions behaves differently for Fe$_{1.141(2)}$Te and
Fe$_{1.076(2)}$Te as a function of temperature, see Fig.~S3. Different structural evolution also reflects in the bond angles.

\begin{figure}

\begin{center}
\includegraphics[width=16cm,angle=0,clip=true]{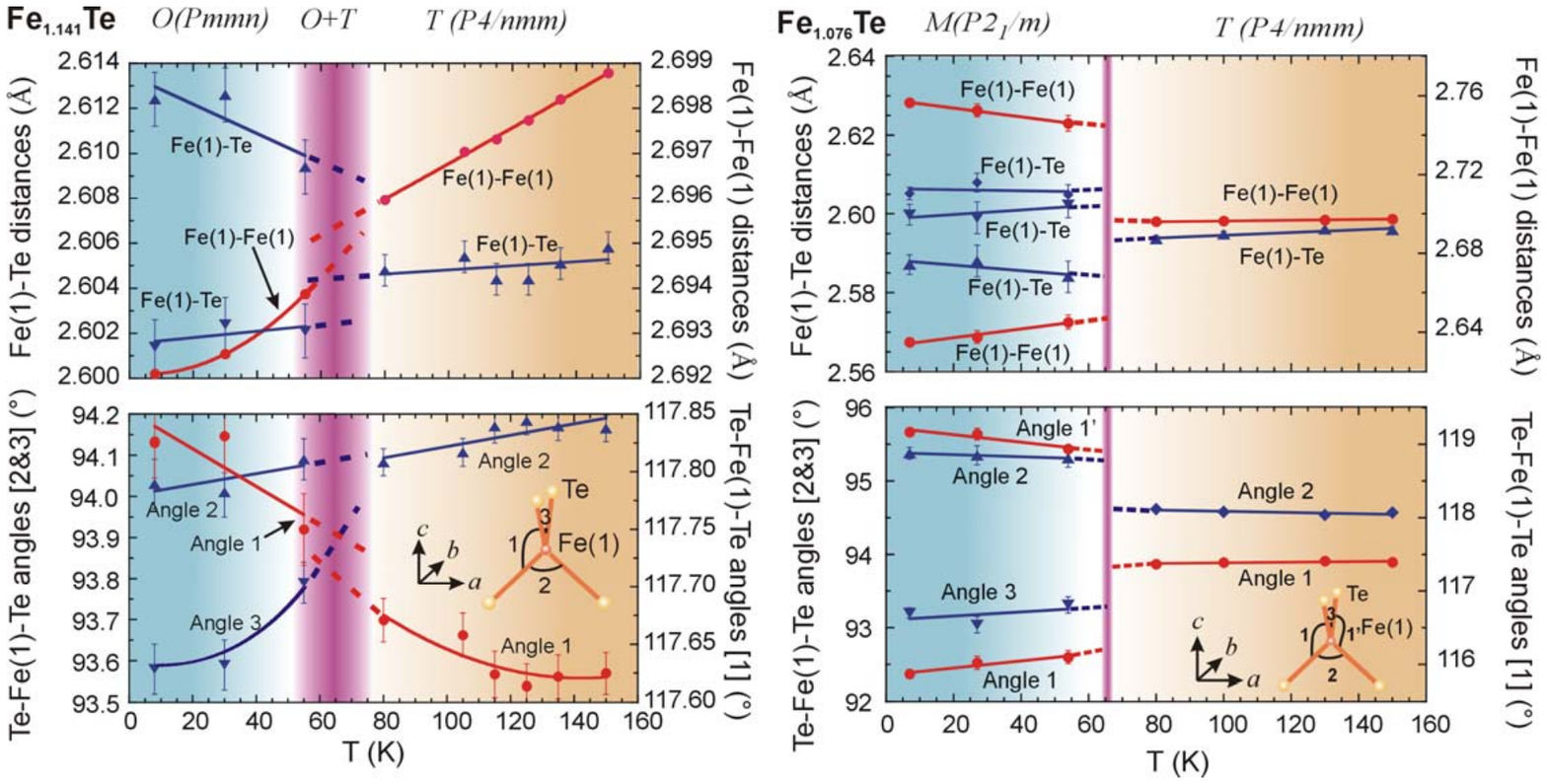}
\end{center}
\vskip -17cm
\caption{{\bf Structure of $\alpha$-FeTe as a function of temperature.}}
\label{figS3}
\end{figure}

\newpage
\includegraphics[scale=.9,angle=0,clip=true]{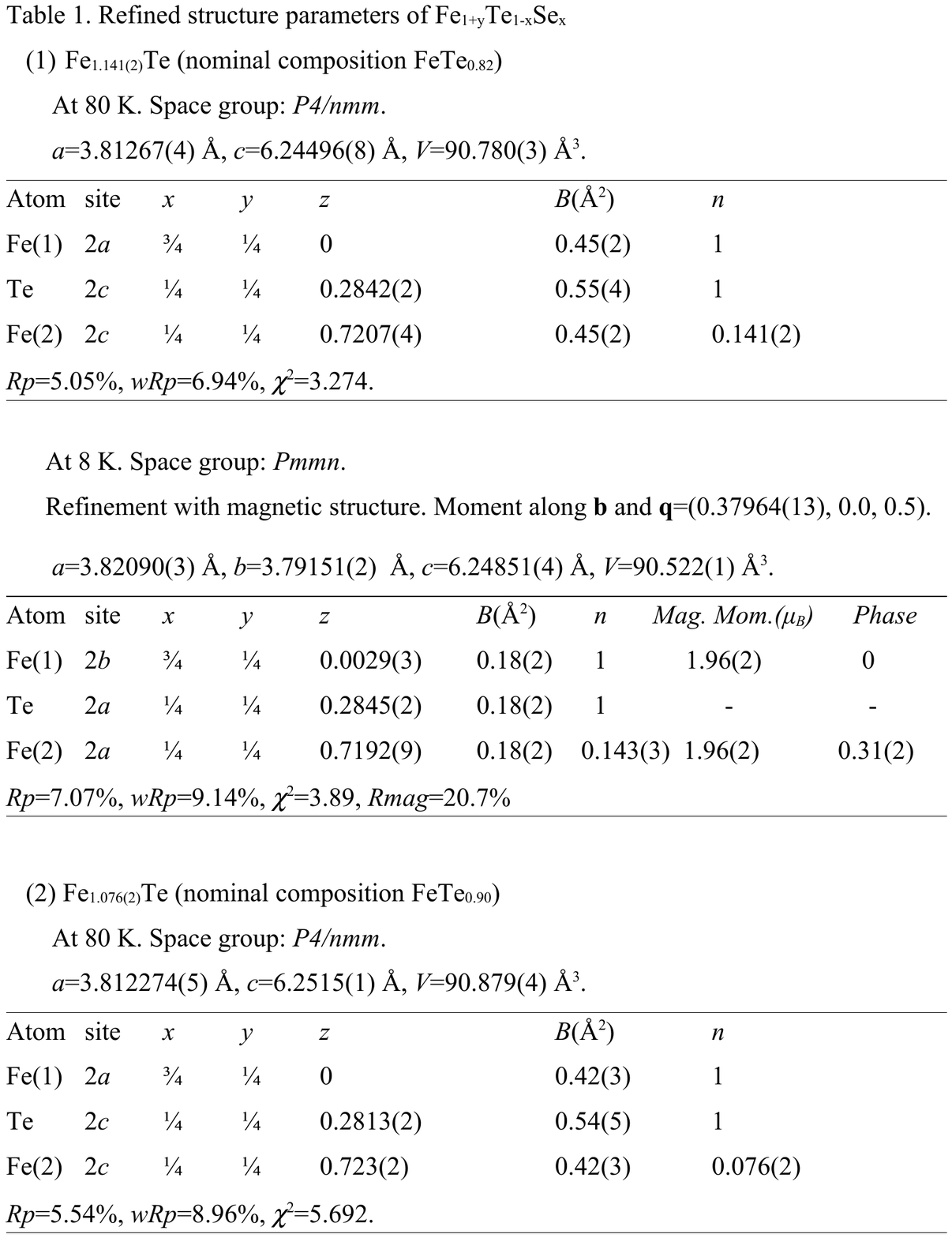}
\newpage
\includegraphics[scale=.9,angle=0,clip=true]{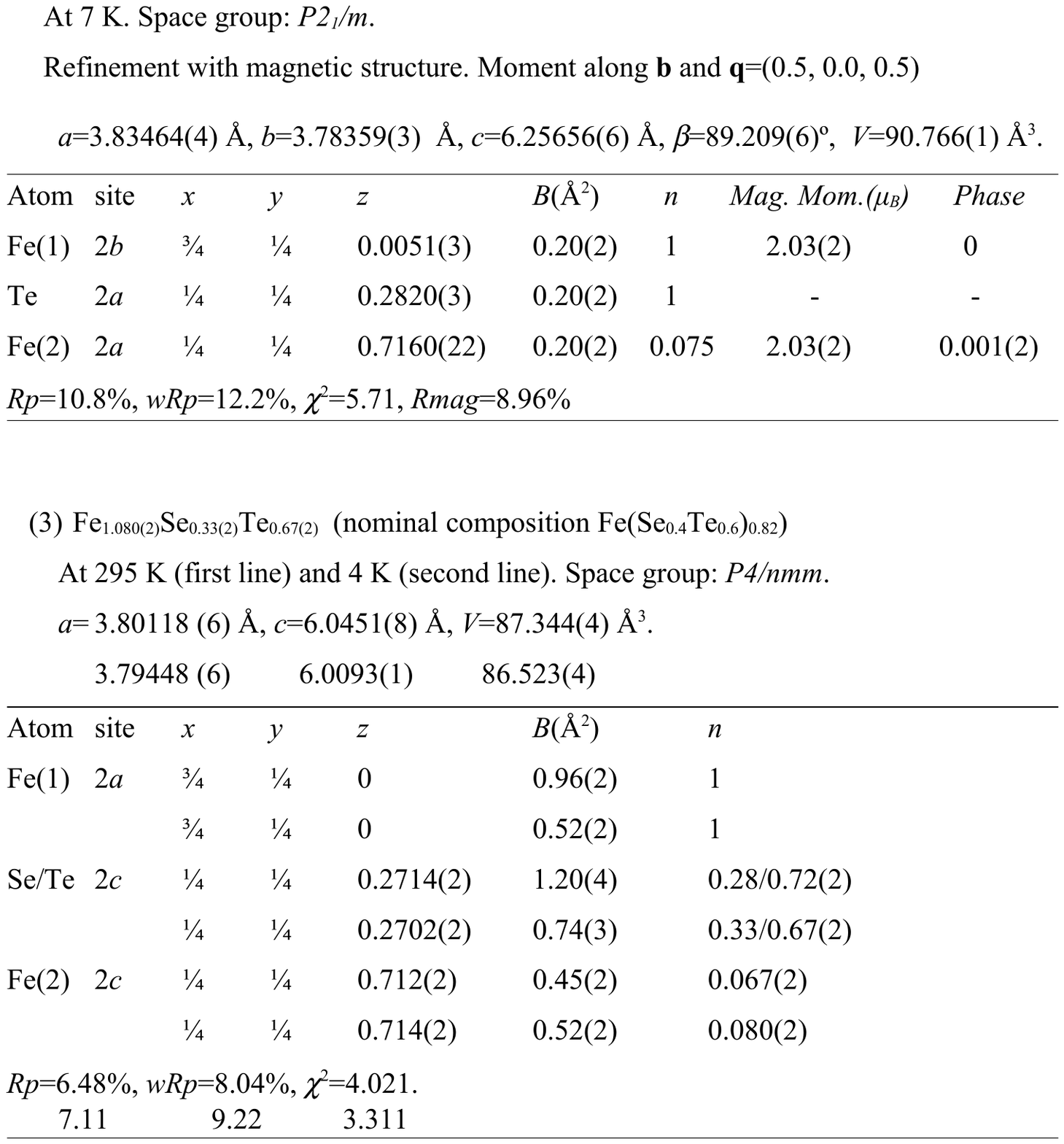}
\newpage
\includegraphics[scale=.9,angle=0,clip=true]{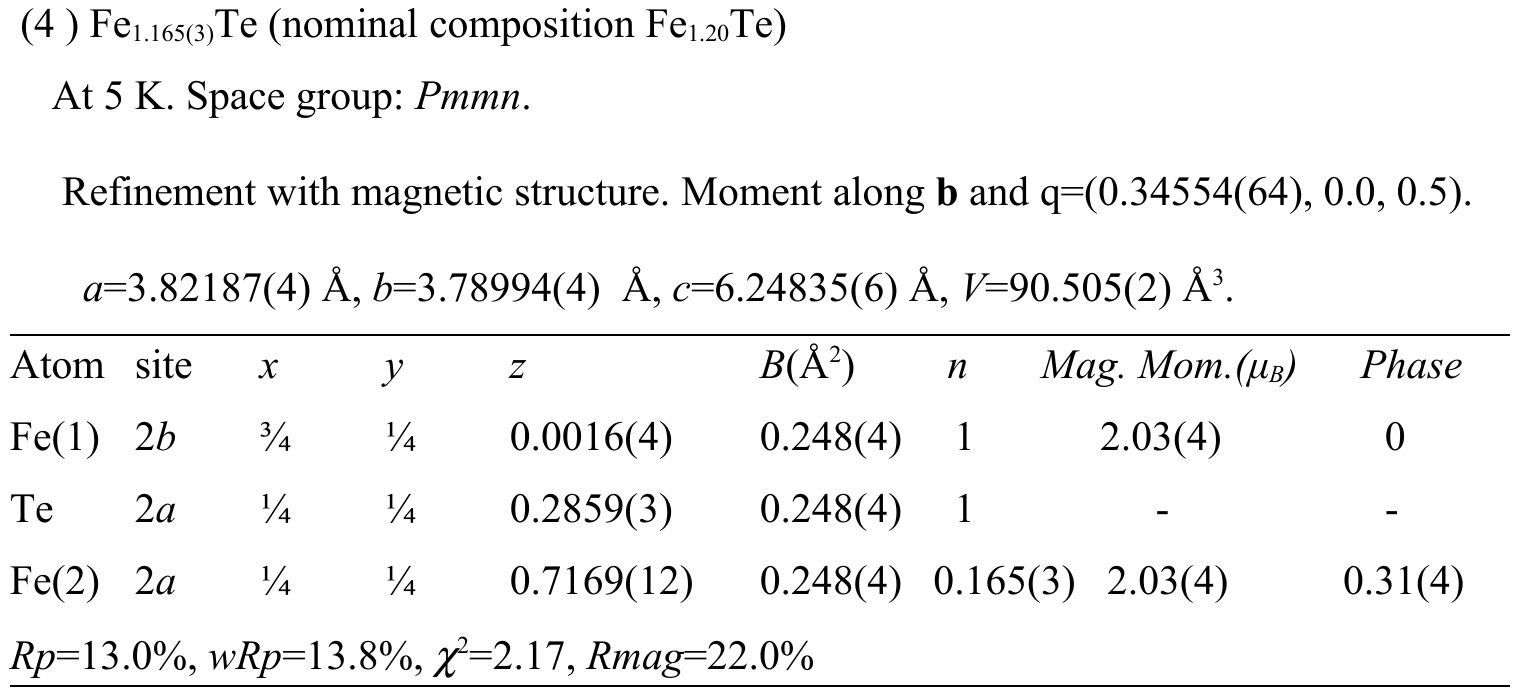}
\end{document}